\documentclass{article}

\usepackage{arxiv}

\usepackage[utf8]{inputenc} % allow utf-8 input
\usepackage[T1]{fontenc}    % use 8-bit T1 fonts
\usepackage{hyperref}       % hyperlinks
\usepackage{url}            % simple URL typesetting
\usepackage{booktabs}       % professional-quality tables
\usepackage{amsfonts}       % blackboard math symbols
\usepackage{nicefrac}       % compact symbols for 1/2, etc.
\usepackage{microtype}      % microtypography
\usepackage{lipsum}		% Can be removed after putting your text content
\usepackage{graphicx}
\usepackage{natbib}
\usepackage{doi}
\usepackage{subfig}

\title{Early Warning Signals for Cryptocurrency Market States}

\author{ \href{https://orcid.org/0000-0002-7691-3784}{\includegraphics[scale=0.06]{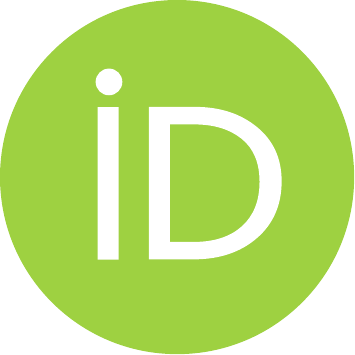}\hspace{1mm}Vishwas Kukreti}\thanks{jai bajrang bali.} \\
	School of Computational \& Integrative Sciences\\
	Jawaharlal Nehru University\\
	New Delhi, 110067 \\
	\texttt{vishwa22\_sit@jnu.ac.in} \\
}

\renewcommand{\shorttitle}{\textit{arXiv} Template}

\hypersetup{
pdftitle={Early Warning Signals for Cryptocurrency Market States},
pdfsubject={q-bio.NC, q-bio.QM},
pdfauthor={Vishwas Kukreti},
pdfkeywords={Cryptocurrency, Graph Kernels, Econophysics},
}

\begin{document}
\maketitle

\begin{abstract}
Being archetypal complex systems, financial markets exhibit rich set of dynamics in their interactions. In this paper, we focus on the recently evolved cryptocurrency market as an example of a complex system and analyse the evolution of cross-correlation structure of cryptocurrencies in the 5 year period from 2017 to 2022. We observe characteristic correlation structures in the observation time window duration and use these specific structures to cluster the cryptocurrency market in 4 market states. %Further, we employ the principal component analysis and show the first component of the correlation matrix acts as an early warning signal for market crash.
\end{abstract}

% keywords can be removed

\section{Introduction}
A financial market is complex system with multiple financial instruments interacting with each other. The change in the behaviour of the market either due to change in economy's structure or macroeconomic episodes such as depression are reflected in the time series of stocks. Time series of assets reflect the overall state of the economy.  Hence, the study of financial time series and their correlation have received substantial interest from researchers. Empirical knowledge states that correlations between asset returns can change significantly from those seen in quieter markets at times of increased market volatility \cite{Sandoval2012, Chakraborti2021,Costa}. As shown by \cite{Munnix2012} the market moves through certain distinct correlation patterns which can be called as market states. Munnix observed correlation between daily returns of stocks that remained in the S\&P500 stocks from 1992-2010 and clustered them using a similarity score which revealed 8 distinct correlation market states in S\&P500. With a slightly improved methodology, \cite{Pharasi2018c} observed 4 market states in S\&P500 in the time period 1985 to 2016.

\section{Data and Methodology}

\subsection{Data description}
The daily closing prices and market capital of 998 cryptocurrencies from 1 January, 2017 to 31 August, 2022 were obtained from CoinGecko, an independent cryptocurrency data aggregator website. CoinGecko employs a global volume-weighted average pricing formula to determine the price of cryptoassets using currency pairings collected from cryptoexchanges. The market capital of an asset is calculated by multiplying its circulating supply with its price. A circulating supply refers to the number of cryptocurrencies that are available in the market to trade. 

\begin{figure}
    \centering
    \includegraphics[scale=0.4]{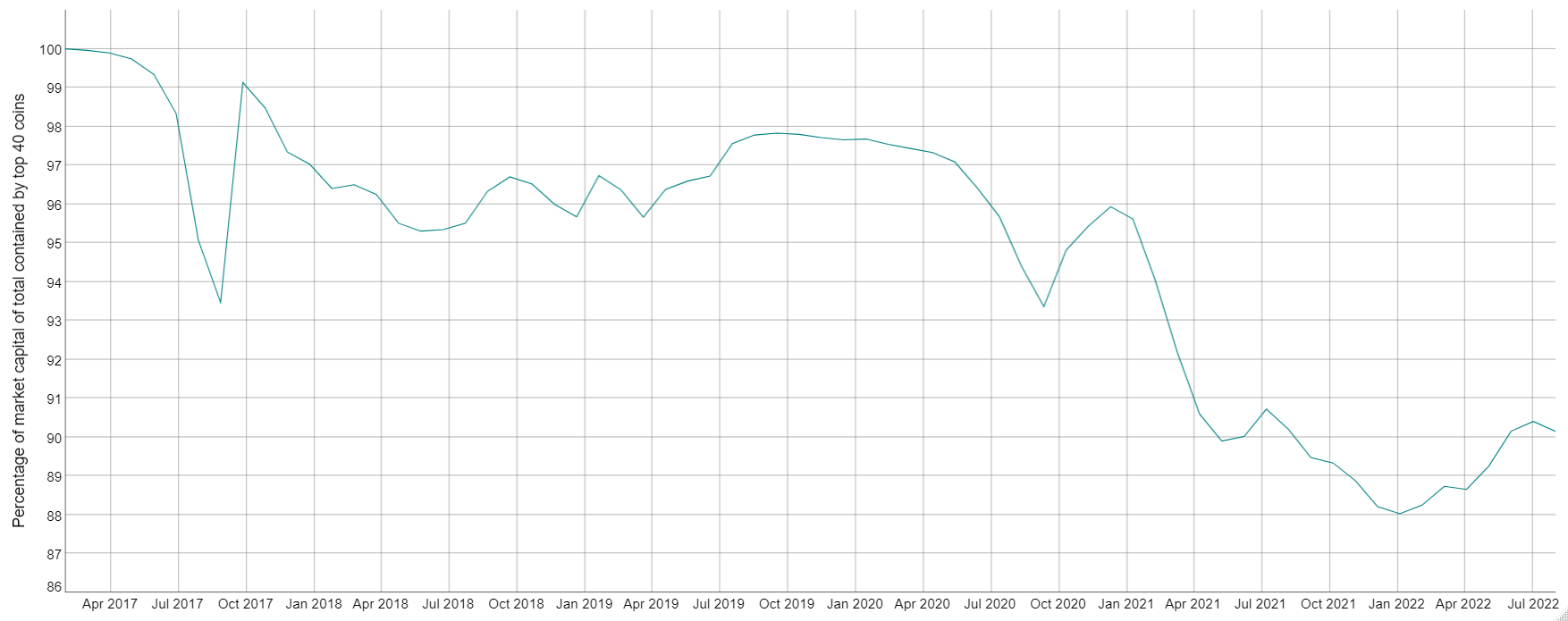}
    \caption{Top 40 coins in every epoch contain the majority of Market Capital in the market. }
    \label{fig:my_label}
\end{figure}

The list of coins is given in Supplementary Information.

\subsection{Methods}

Munnix et. al. consider only the stocks present in the entire duration between 1992-2010 in S\&P 500 to create correlation matrix. Such methodology skips on stocks which could have been important before they were wiped out of the market. This approach excludes stocks like Lehman Brothers (dissolved after sub-prime mortgage crash of 2008), Pets.com (shut down after dot-com bubble bust) etc. which had a strong market presence but could not survive the market turbulence. Such approach makes analysis of cryptocurrencies more difficult where on an average 7 currencies appear and disappear from the market every week \cite{ElBahrawy}. Hence, we proceed with a portfolio approach where every epoch consists of top 40 coins in that epoch. This solves another problem of stationarity in time series while calculating correlation. When $T/K < 1$, where $T$ is the time horizon over which correlation is calculated and $K$ is the number of stocks, the correlation matrix becomes singular. For our particular case, majority of the market dynamics can be evaluated using top 40 market participants in an time horizon of 60 days. However, the change in top 40 tokens in every epoch requires an new similarity measure than used in earlier literature. Hence, we convert the correlation matrices to graphs and use graph kernels to calculate a similarity measure between the matrices.

In this study, we analyse the temporal evolution of the cross-correlation patterns over different epochs in return time series for cryptocurrencies and identify the ideal number of market states. The daily return time series is constructed as $r_{n}(t)=\ln S_{n}(t)-\ln S_{n}(t-\ 1)$, where $S_{n}$ is the closing price of stock $n$ at time $t$. The cross-correlation matrix is constructed using Pearson cross-correlation coefficients, $C_{i j}(\tau)=(\langle r_{i}r_{j}\rangle\,-\,\langle r_{i}\rangle\,\langle r_{j}\rangle)/\sigma_{i}\sigma_{j},$, where $i,j=\,1,...,N,\tau$ denote the epoch of length $N$ days. 

\begin{figure}[h]
    \centering
    \includegraphics[scale=0.3]{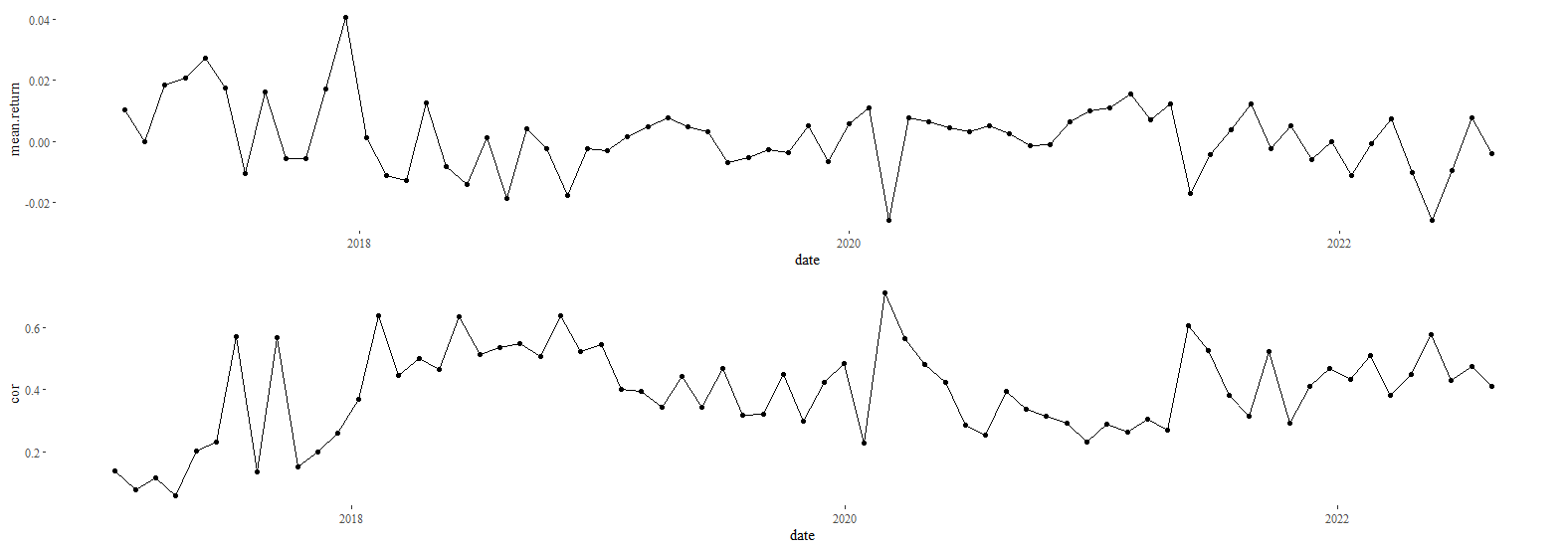}
     \caption{Evolution of market correlation and return. Top row shows mean return of the 40 coin portfolio per epoch. Bottom row shows the mean correlation. Whenever mean return falls, correlation increases in the market.  }
    \label{mean_ret_mean_cor}
\end{figure}

Figure \ref{mean_ret_mean_cor} displays the evolution of the mean return of the 40 coin portfolio per epoch and mean market correlation. Notably, anytime a market crash occurs the mean market correlation rises significantly, showing that the market is highly correlated and the majority of the stocks behave in a similar fashion.

\begin{figure}[h]
    \centering
    \includegraphics[scale=0.3]{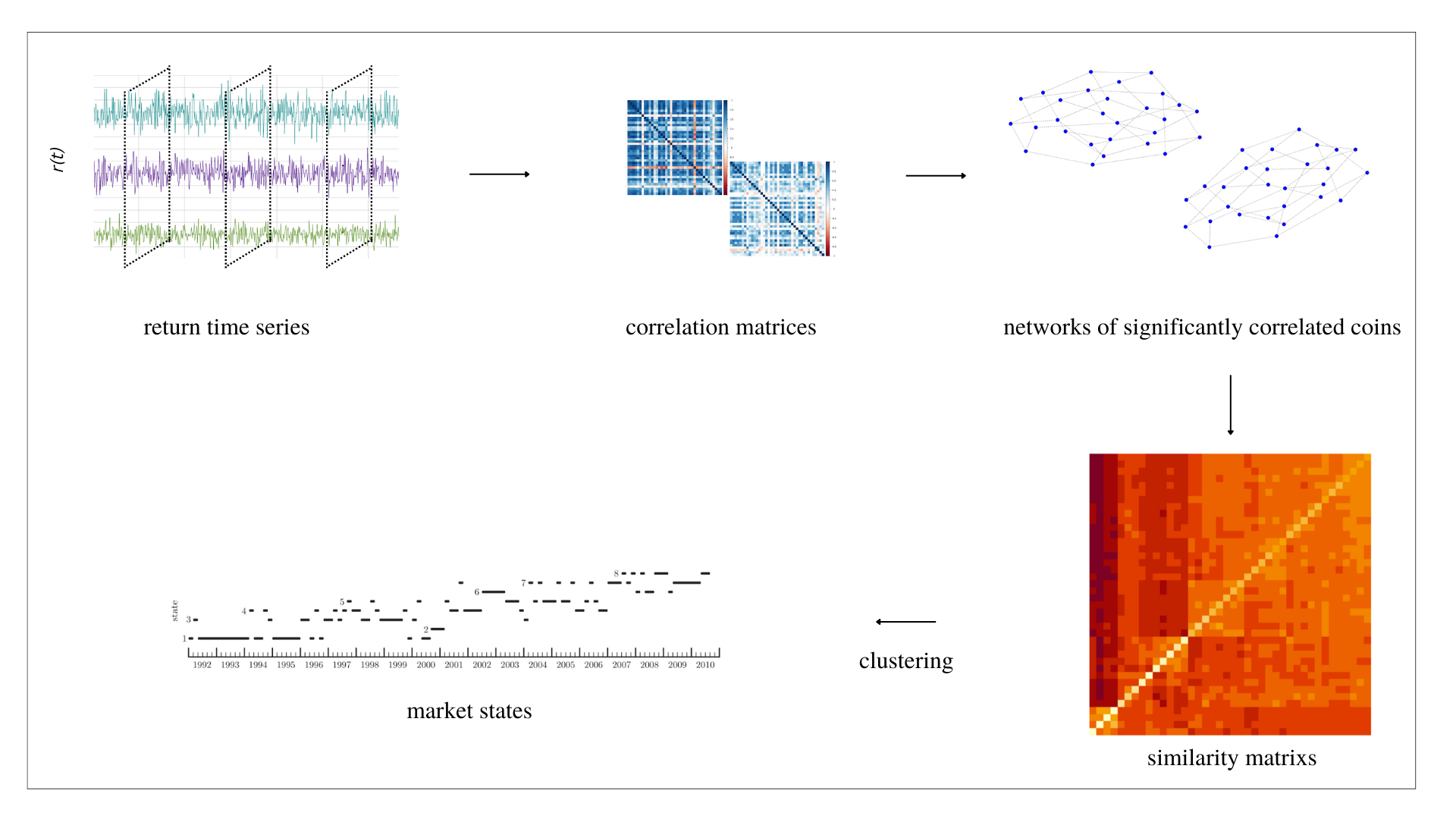}
    \caption{Methodology}
    \label{methodsFig}
\end{figure}

\subsubsection{Creating network from correlation matrices}

We consider short disjoint time windows ($T$) of 20 days to calculate the statistical dependence between cryptocurrencies. For every epoch, top 40 coins by market capital were selected to create the correlation matrix. The empirical returns of crytpocurrencies are not normal distributed, rather fat tailed, and their drift, volatility and, autocorrelation cannot be assumed constant. Since the stationarity assumption fails for longer time series, we divide the return series into shorter epochs and treat it for local stationarity. For $n$ most recent data points in return series, the locally normalized return series is constructed as:
$\bar{r}(t)\equiv\frac{r\langle t\rangle-\langle r(t)\rangle_{n}}{\sqrt{\langle r^{2}(t)\rangle_{n}-\langle r(t)\rangle_{n}^2}}$
As suggested by authors in \cite{guhr2003new}, the value of $n$ was fixed to 13.

The shorter epochs lead to singular correlation matrices i.e. when $T/K < 1$, where $T$ is the time horizon over which correlation is calculated and $K$ is the number of assets, the correlation matrix becomes singular. As \cite{doi:10.1080/14697688.2010.534813} note, the benefit of covariance matrix estimators strongly relies on the ratio between the estimation period $T$ and the number of stocks $N$. To deal with the singularity of matrices, we treat the cross-correlation matrices with power map technique introduced by \cite{guhr2003new} to suppress the noise. In power map technique, every element of the cross-correlation matrix is given a non-linear distortion in the form of: $\left.C_{i j}\rightarrow_{}\right.(\mathrm{sign}\mathrm{C_{i j}})\left|C_{i j}\right|^{q}$, where $q$ is the noise-suppression parameter. We choose $q=1.5$ for two reasons. For $q > 1.5$ the network created from correlation matrices, especially during crash periods, became empty leading to failure of graph similarity algorithms. Secondly, previous literature recommends using $q=1.5$ for optimal noise to signal ratio.

Since the seminal paper of Mantegna in 1999 \cite{Mantegna1999}, various algorithms and methodologies like Minimum Spanning Tree (MST), Planar Maximally Filtered Graph (PMFG) etc. are available to study the evolving taxonomy of assets. Most of these methods convert correlation matrix to distance matrix and optimise the distance matrix to an adjacency matrix as an input for network \cite{Martib}. Here, we utilise a new and novel methodology, \textit{Scola}, presented in \cite{doi:10.1098/rspa.2019.0578} to convert correlation matrices to graph. If the correlation between nodes can be explained by random fluctuations under a white noise null model, which presupposes that observations at all nodes are independent of one another, then Scola considers the correlation between nodes to be spurious. However, If a chosen null model cannot explain the correlations, then Scola adds an edge between node pairings. 

\begin{figure}[h]
    \centering
    \includegraphics[scale=0.1]{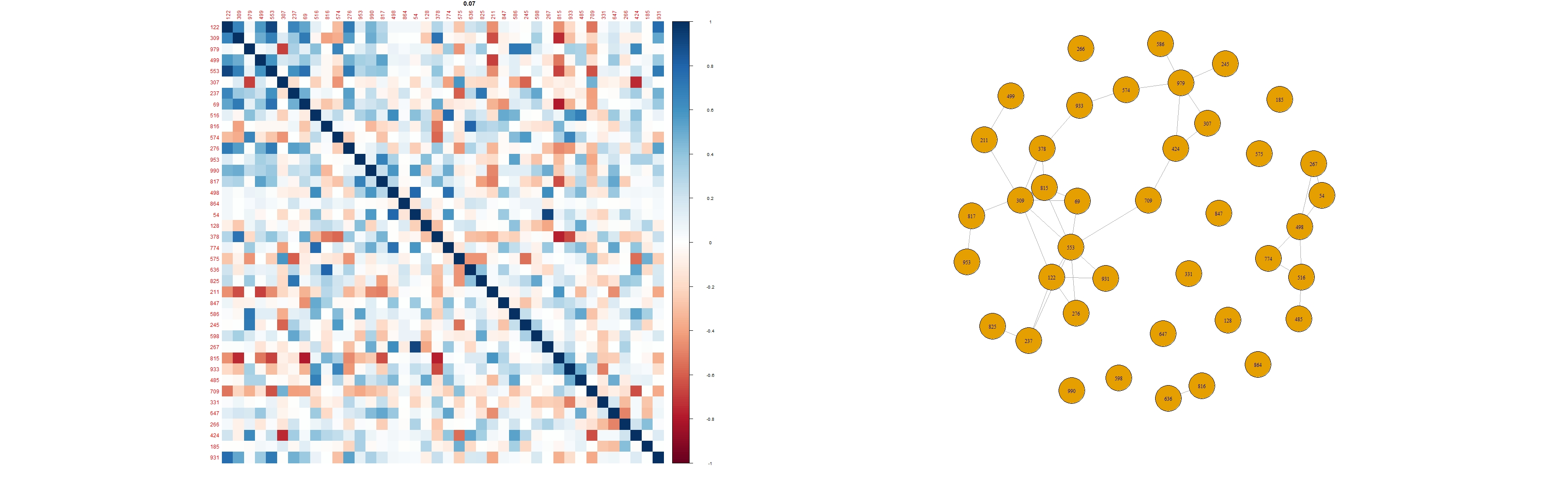}
     \includegraphics[scale=0.1]{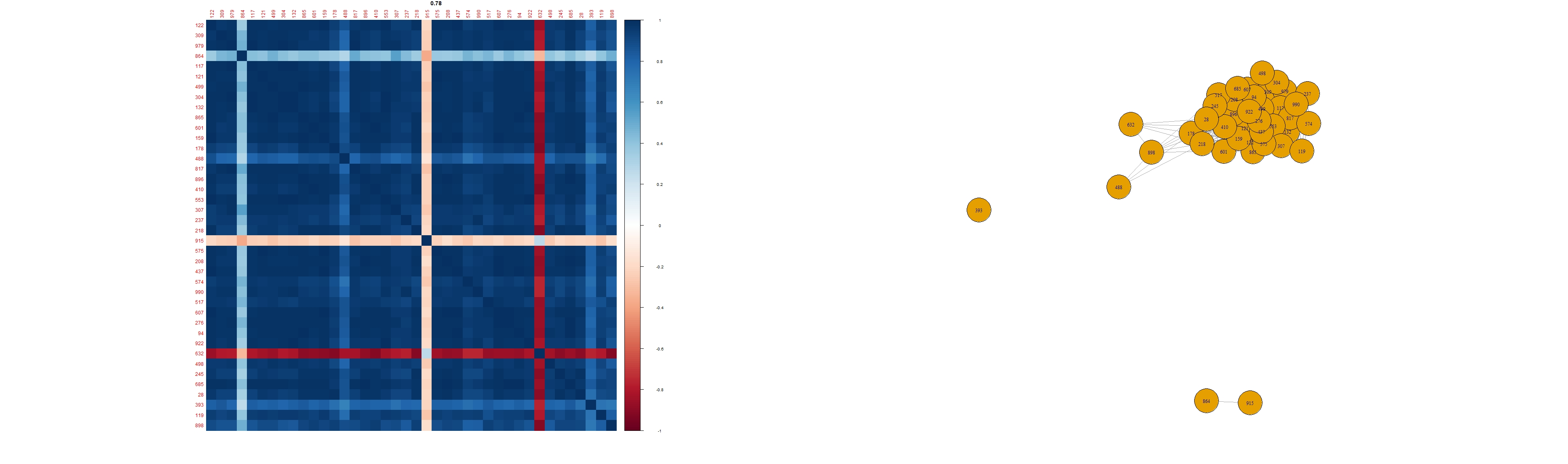}
    \caption{Left figure corresponds to market order on 2017-01-10, normal market duration. Right figure corresponds to 2020-03-15, market crash due to Covid scare. When correlation between currencies increases, crypto graph clustering increases. }
    \label{fig:my_label}
\end{figure}

\subsubsection{Calculating similarity between asset graphs using graph kernels}

A kernel $K(x,{x}')$ is a measure of similarity between two graphs $x$ and ${x}'$. A graph kernel is a symmetric, i.e. $k(x,{x}') = k({x}',x)$, and a positive semidefinite function defined on the space of graphs $G$. By defining a distance function $d:G \times G \rightarrow  \mathbb{R}+$ we can compare two graphs. When comparing nodes in a graph, a kernel between the nodes is calculated, however when comparing graphs, a kernel between graphs is calculated. Numerous graph similarity metrics are based on graph isomorphism or related ideas like the largest common subgraph or subgraph isomorphism. Checking whether two graphs are topologically identical, or isomorphic, is perhaps the most straightforward way to determine how similar they are. To cluster the 103 asset networks obtained, we use graph kernels based on Weisfeiler-Lehman isomorphism test \cite{Shervashidze2011,Weisfeiler1968}. The key idea of the algorithm is to augment the node labels by the sorted set of node labels of neighbouring nodes, and compress these augmented labels into new, short labels. Then, these actions are repeated until the node label sets of $G$ and $G^{\prime}$ differ or n iterations have been performed. After $n$ iterations, if the sets are identical, either $G$ and $G^{\prime}$ are isomorphic or the algorithm was unable to detect that they are not.

\subsubsection{Clustering networks into market states}
To determing the number of clusters \textit{i.e} market states, we emply eigen gap heuristics. According to the eigengap heuristic, the value of $k$ that maximises the eigengap often represents the number of clusters (difference between consecutive eigenvalues). The closer the eigenvectors are in the ideal scenario, the better spectral clustering performs, hence the wider this eigengap is. The networks obtained were clustered using k-means clustering algorithm. 

\begin{figure}[h]
    \centering
    \includegraphics[scale=0.3]{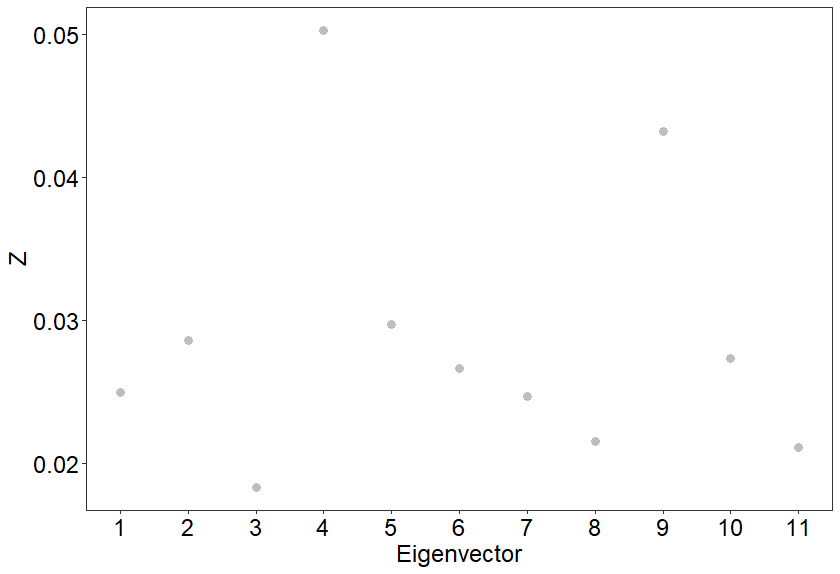}
    \includegraphics[scale=0.3]{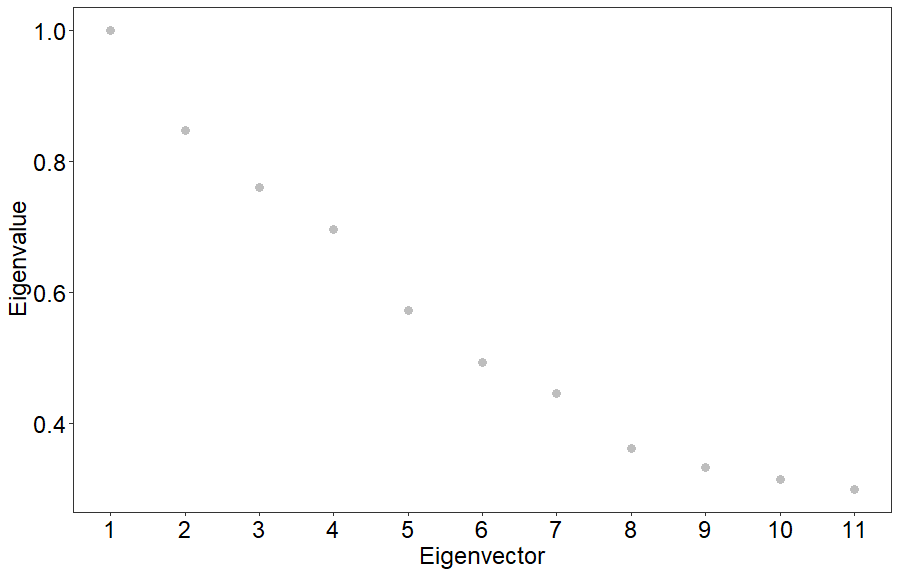}
    \caption{Largest eigen gap appears at 4.}
    \label{fig:my_label}
\end{figure}

\begin{figure}[h]
    \centering
    \includegraphics[scale=0.3]{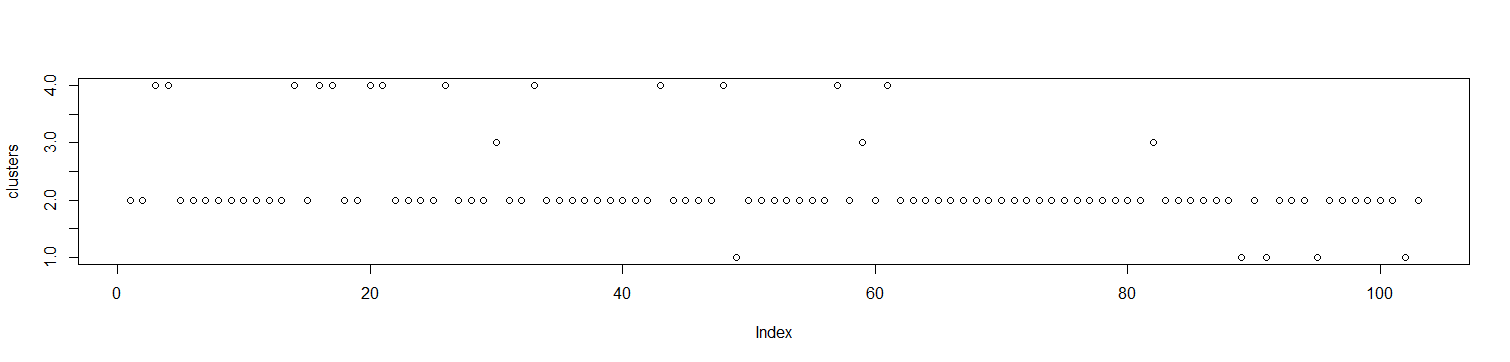}
    \caption{Clusters}
    \label{fig:my_label}
\end{figure}

\section{Results \& Discussions}

Based on k-means clustering, the 103 correlation matrices were clustered in 4 distance market states. The four clusters refer to the state of cryptomarket at different dates from 2017 to 2022.

\begin{figure}[h]
    \centering

\subfloat[Market State 1]{\includegraphics[width=0.25\textwidth]{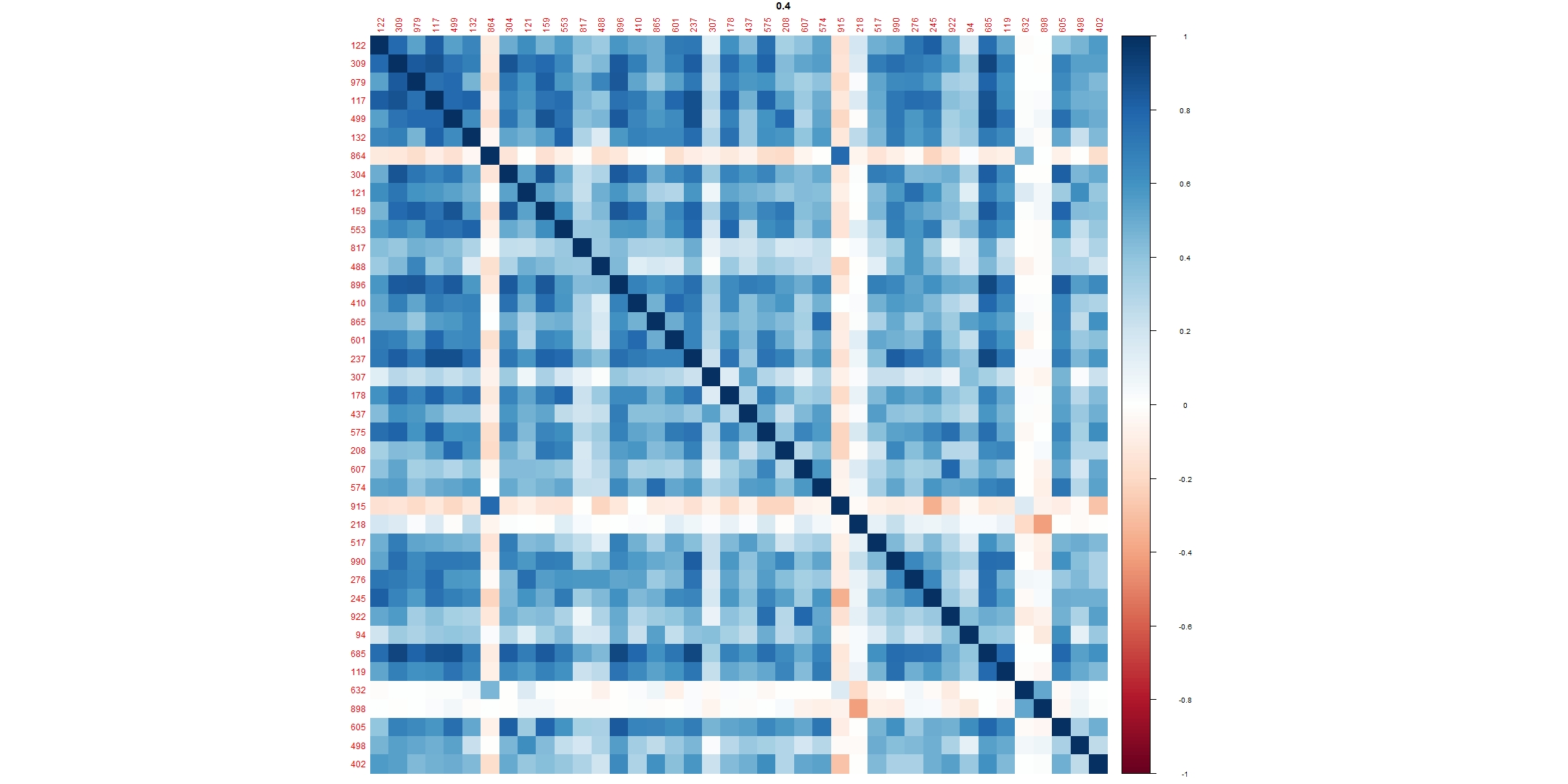}}
\subfloat[Market State 2]{\includegraphics[width=0.25\textwidth]{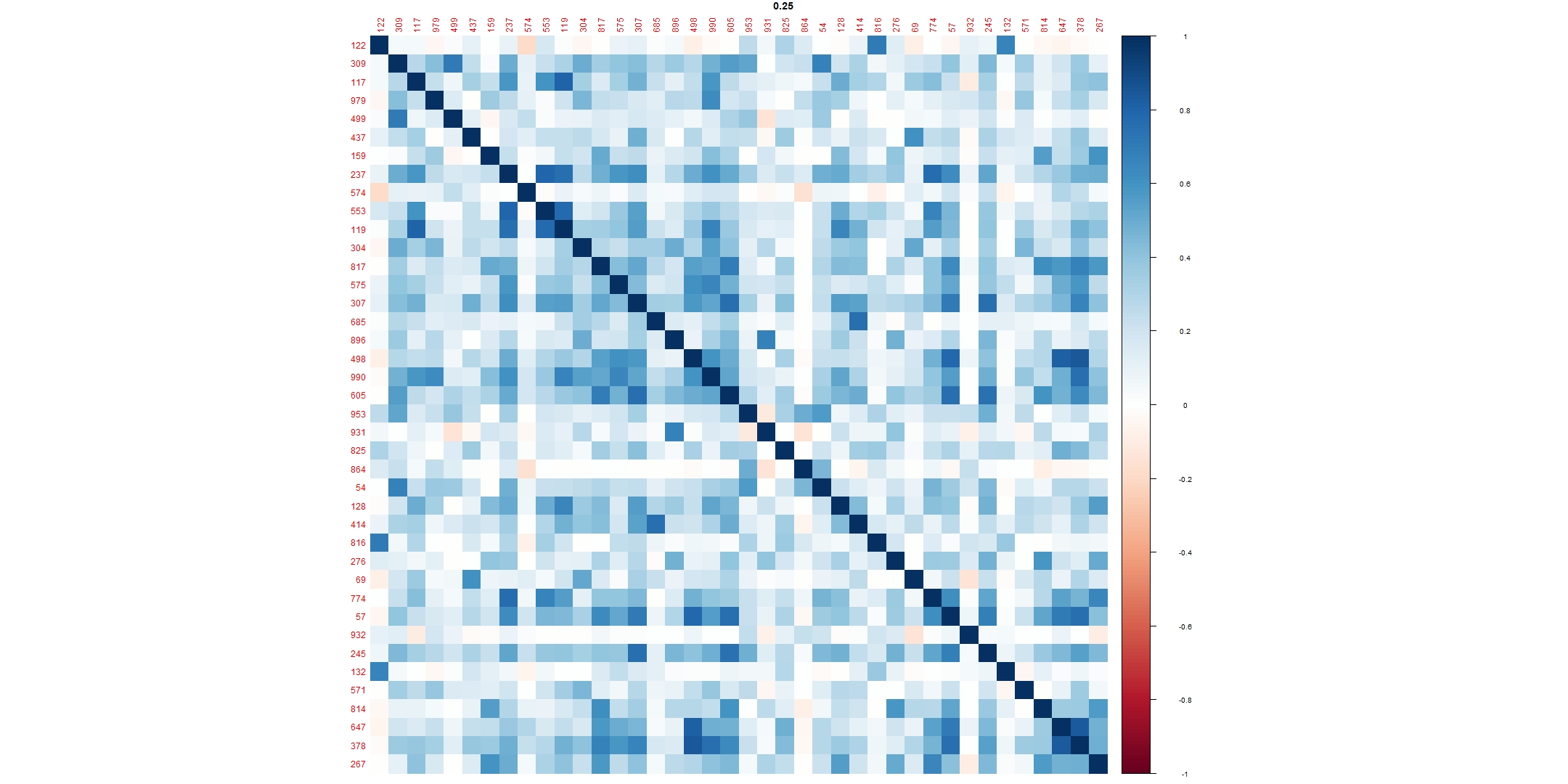}}
\subfloat[Market State 3]{\includegraphics[width=0.25\textwidth]{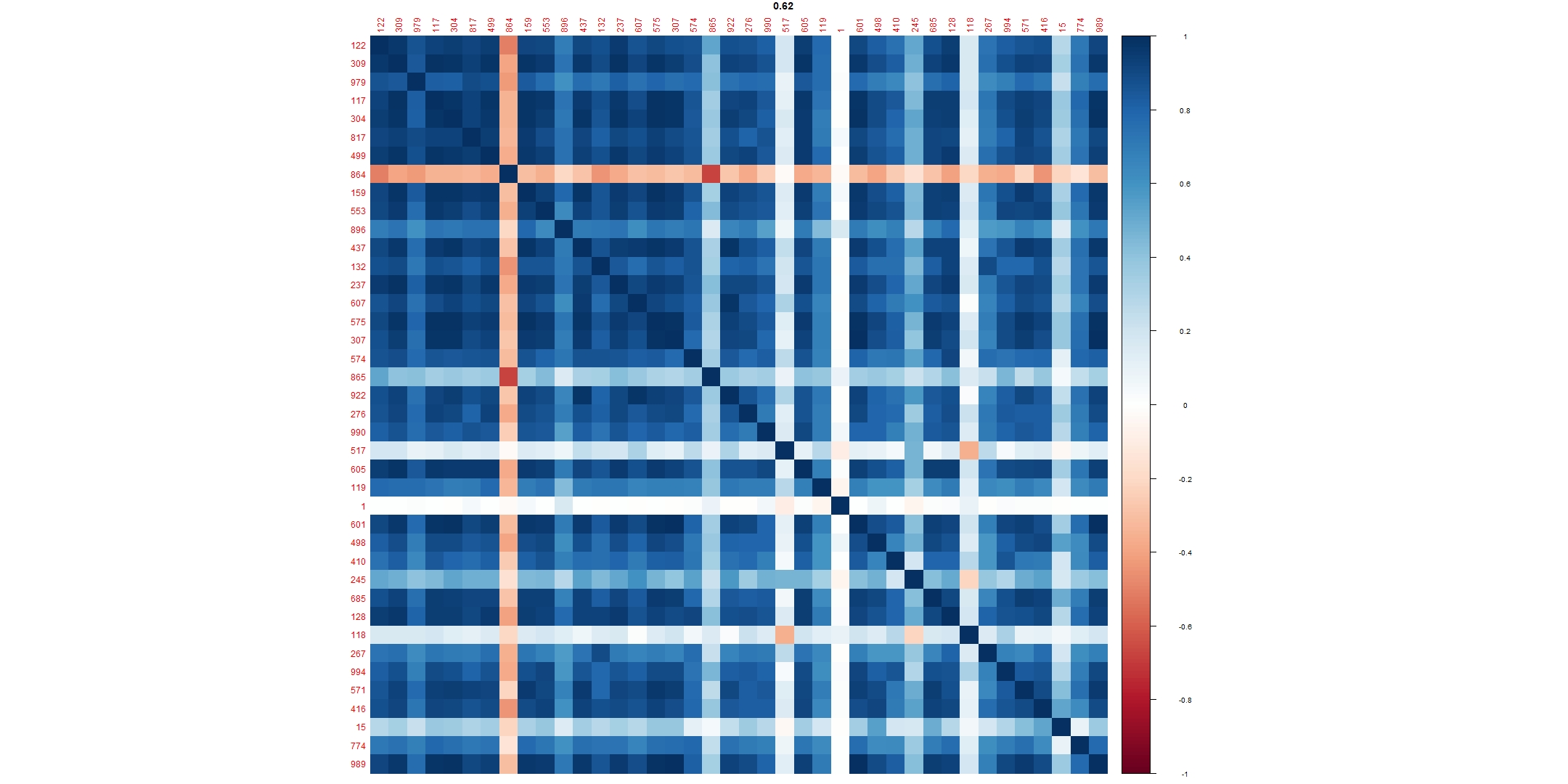}}
\subfloat[Market State 4]{\includegraphics[width=0.25\textwidth]{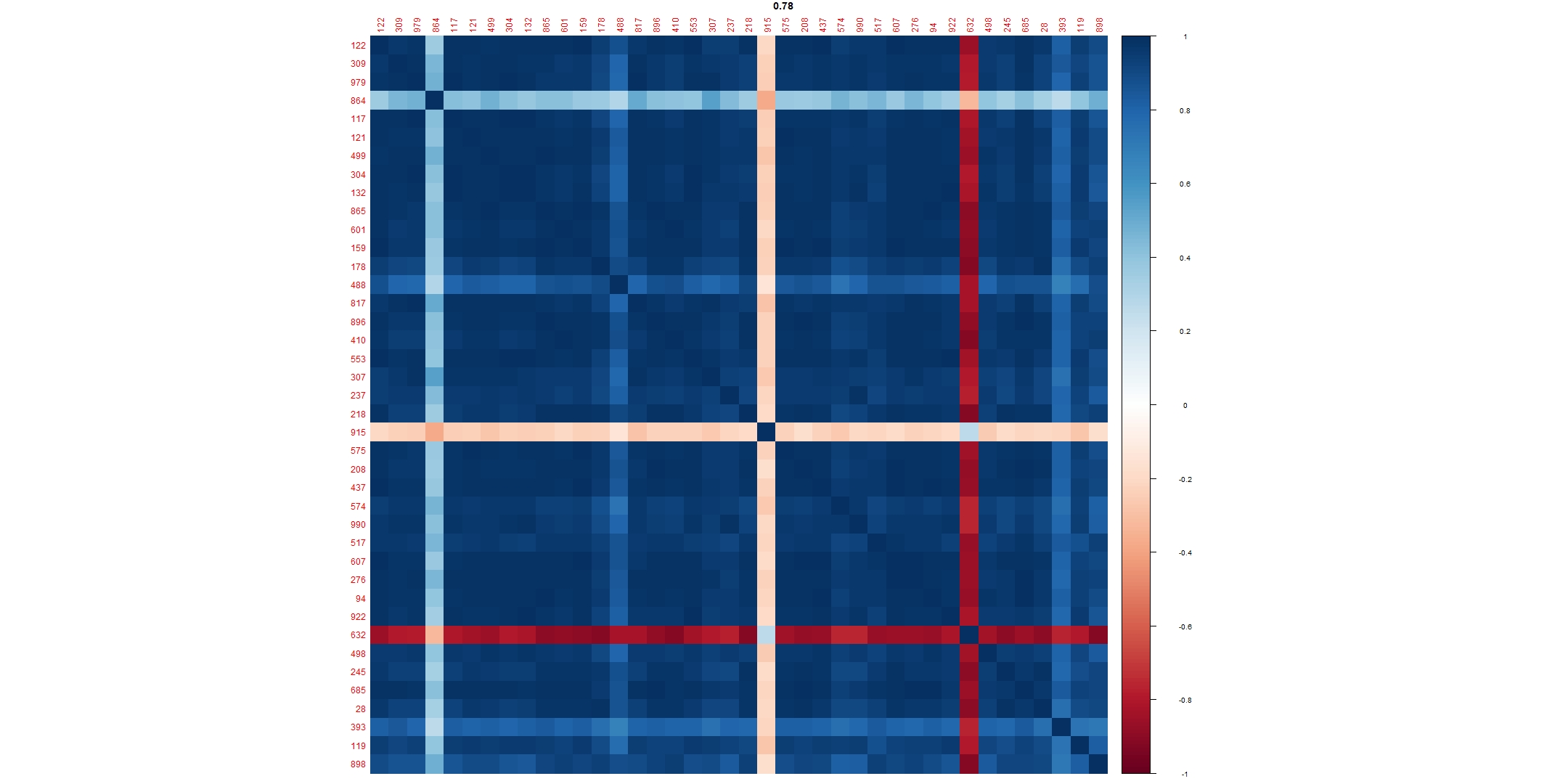}}
\caption{Market States}
\end{figure}

\section{Further works}

\begin{enumerate}
    \item A parameter for assessing the quality of clusters is not utilised as ground truth is not available.
\end{enumerate}

\bibliographystyle{plainnat}
\bibliography{references} 

\begin{thebibliography}{13}
\providecommand{\natexlab}[1]{#1}
\providecommand{\url}[1]{\texttt{#1}}
\expandafter\ifx\csname urlstyle\endcsname\relax
  \providecommand{\doi}[1]{doi: #1}\else
  \providecommand{\doi}{doi: \begingroup \urlstyle{rm}\Url}\fi

\bibitem[Chakraborti et~al.(2021)Chakraborti, Hrishidev, Sharma, and
  Pharasi]{Chakraborti2021}
Anirban Chakraborti, Hrishidev, Kiran Sharma, and Hirdesh~K. Pharasi.
\newblock {Phase separation and scaling in correlation structures of financial
  markets}.
\newblock \emph{Journal of Physics: Complexity}, 2\penalty0 (1), mar 2021.
\newblock ISSN 2632072X.
\newblock \doi{10.1088/2632-072X/abbed1}.

\bibitem[Costa et~al.(2011)Costa, Jr, in~{\ldots}, and undefined 2011]{Costa}
LF~Costa, ON~Oliveira Jr, G~Travieso~Advances in~{\ldots}, and undefined 2011.
\newblock {Analyzing and modeling real-world phenomena with complex networks: a
  survey of applications}.
\newblock \emph{Taylor {\&} Francis}, 2011.
\newblock URL
  \url{https://www.tandfonline.com/doi/abs/10.1080/00018732.2011.572452}.

\bibitem[ElBahrawy et~al.(2017)ElBahrawy, Alessandretti, Kandler,
  Pastor-Satorras, and Baronchelli]{ElBahrawy}
Abeer ElBahrawy, Laura Alessandretti, Anne Kandler, Romualdo Pastor-Satorras,
  and Andrea Baronchelli.
\newblock Evolutionary dynamics of the cryptocurrency market.
\newblock \emph{Royal Society Open Science}, 4\penalty0 (11):\penalty0 170623,
  2017.
\newblock \doi{10.1098/rsos.170623}.
\newblock URL
  \url{https://royalsocietypublishing.org/doi/abs/10.1098/rsos.170623}.

\bibitem[Guhr and K{\"a}lber(2003)]{guhr2003new}
Thomas Guhr and Bernd K{\"a}lber.
\newblock A new method to estimate the noise in financial correlation matrices.
\newblock \emph{Journal of Physics A: Mathematical and General}, 36\penalty0
  (12):\penalty0 3009, 2003.

\bibitem[Kojaku and Masuda(2019)]{doi:10.1098/rspa.2019.0578}
Sadamori Kojaku and Naoki Masuda.
\newblock Constructing networks by filtering correlation matrices: a null model
  approach.
\newblock \emph{Proceedings of the Royal Society A: Mathematical, Physical and
  Engineering Sciences}, 475\penalty0 (2231):\penalty0 20190578, 2019.
\newblock \doi{10.1098/rspa.2019.0578}.
\newblock URL
  \url{https://royalsocietypublishing.org/doi/abs/10.1098/rspa.2019.0578}.

\bibitem[Mantegna(1999)]{Mantegna1999}
R.N. Mantegna.
\newblock {Hierarchical structure in financial markets}.
\newblock \emph{The European Physical Journal B}, 11\penalty0 (1):\penalty0
  193--197, sep 1999.
\newblock ISSN 1434-6028.
\newblock \doi{10.1007/s100510050929}.
\newblock URL \url{http://link.springer.com/10.1007/s100510050929}.

\bibitem[Marti et~al.(2017)Marti, Nielsen, Bi{\'{n}}kowski, and Donnat]{Martib}
Gautier Marti, Frank Nielsen, Miko{\l}aj Bi{\'{n}}kowski, and Philippe Donnat.
\newblock {A review of two decades of correlations, hierarchies, networks and
  clustering in financial markets}.
\newblock \emph{arxiv.org}, mar 2017.
\newblock \doi{10.1007/978-3-030-65459-7}.
\newblock URL \url{https://arxiv.org/abs/1703.00485
  http://arxiv.org/abs/1703.00485 http://dx.doi.org/10.1007/978-3-030-65459-7}.

\bibitem[M{\"{u}}nnix et~al.(2012)M{\"{u}}nnix, Shimada, Sch{\"{a}}fer,
  Leyvraz, Seligman, Guhr, and Stanley]{Munnix2012}
Michael~C. M{\"{u}}nnix, Takashi Shimada, Rudi Sch{\"{a}}fer, Francois Leyvraz,
  Thomas~H. Seligman, Thomas Guhr, and H.~Eugene Stanley.
\newblock {Identifying states of a financial market}.
\newblock \emph{Scientific Reports}, 2:\penalty0 1--6, 2012.
\newblock ISSN 20452322.
\newblock \doi{10.1038/srep00644}.

\bibitem[Pantaleo et~al.(2011)Pantaleo, Tumminello, Lillo, and
  Mantegna]{doi:10.1080/14697688.2010.534813}
Ester Pantaleo, Michele Tumminello, Fabrizio Lillo, and Rosario~N. Mantegna.
\newblock When do improved covariance matrix estimators enhance portfolio
  optimization? an empirical comparative study of nine estimators.
\newblock \emph{Quantitative Finance}, 11\penalty0 (7):\penalty0 1067--1080,
  2011.
\newblock \doi{10.1080/14697688.2010.534813}.
\newblock URL \url{https://doi.org/10.1080/14697688.2010.534813}.

\bibitem[Pharasi et~al.(2018)Pharasi, Sharma, Chatterjee, Chakraborti, Leyvraz,
  and Seligman]{Pharasi2018c}
Hirdesh~K. Pharasi, Kiran Sharma, Rakesh Chatterjee, Anirban Chakraborti,
  Francois Leyvraz, and Thomas~H. Seligman.
\newblock {Identifying long-term precursors of financial market crashes using
  correlation patterns}.
\newblock \emph{New Journal of Physics}, 20\penalty0 (10):\penalty0 103041,
  2018.
\newblock ISSN 13672630.
\newblock \doi{10.1088/1367-2630/aae7e0}.
\newblock URL \url{https://doi.org/10.1088/1367-2630/aae7e0}.

\bibitem[Sandoval and Franca(2012)]{Sandoval2012}
Leonidas Sandoval and Italo De~Paula Franca.
\newblock {Correlation of financial markets in times of crisis}.
\newblock \emph{Physica A: Statistical Mechanics and its Applications},
  391\penalty0 (1-2):\penalty0 187--208, 2012.
\newblock ISSN 03784371.
\newblock \doi{10.1016/j.physa.2011.07.023}.
\newblock URL \url{http://dx.doi.org/10.1016/j.physa.2011.07.023}.

\bibitem[Shervashidze et~al.(2011)Shervashidze, Schweitzer, {Van Leeuwen},
  Mehlhorn, and Borgwardt]{Shervashidze2011}
Nino Shervashidze, Pascal Schweitzer, Erik~Jan {Van Leeuwen}, Kurt Mehlhorn,
  and Karsten~M. Borgwardt.
\newblock {Weisfeiler-Lehman graph kernels}.
\newblock \emph{Journal of Machine Learning Research}, 12:\penalty0 2539--2561,
  2011.
\newblock ISSN 15337928.

\bibitem[Weisfeiler and Leman(1968)]{Weisfeiler1968}
B~Yu Weisfeiler and A~A Leman.
\newblock A reduction of a graph to a canonical form and an algebra arising
  from this reduction.
\newblock \emph{Nauchno-Technicheskaya Informatsia}, 2, 1968.

\end{thebibliography}

\end{document}